\documentclass[10pt,letterpaper,twocolumn]{article} 

\usepackage{ol2}
\usepackage[draft]{hyperref}
\usepackage{amsmath}

\begin{document}

\twocolumn[ 

\title{Temperature-dependent bulk viscosity of nitrogen gas determined from spontaneous Rayleigh-Brillouin scattering}


\author{Ziyu Gu, Wim Ubachs$*$}

\address{Department of Physics and Astronomy, LaserLaB, VU University,
\\ De Boelelaan 1081, 1081 HV Amsterdam, The Netherlands
\\

$^*$Corresponding author: w.m.g.ubachs@vu.nl
}

\begin{abstract} Values for the bulk viscosity $\eta_b$ of molecular nitrogen gas (N$_2$) were derived from spontaneous Rayleigh-Brillouin (RB) scattering at ultraviolet wavelengths ($\lambda= 366.8$ nm) and at a $90^\circ$ scattering angle. Analysis of the scattering profiles yield values showing a linear increasing trend, ranging from $\eta_b = 0.7 \times 10^{-5}$ to $2.0 \times 10^{-5}$ kg$\cdot$m$^{-1}$$\cdot$s$^{-1}$ in the temperature interval from 255 K to 340 K. The present values, pertaining to hypersound acoustics at frequencies in the GHz domain, are found to be in agreement with results from acoustic attenuation experiments in N$_2$ performed at MHz frequencies.

\end{abstract}


 ] 

The concept of bulk viscosity, $\eta_b$, also referred to as volume viscosity is part of a thermodynamic description of gases as a transport coefficient in addition to the shear viscosity $\eta_s$~\cite{Landau1959,Morse1986}. The bulk viscosity results from the collisional energy exchange between the translational and internal (rotational and vibrational) degrees of freedom in fluids. The value of $\eta_b$ of gases can be measured via sound absorption but only a limited number of studies have been reported~\cite{Kneser1933,Prangsma1973}. Furthermore, such measurements yield values for $\eta_b$ related to acoustic frequencies in the MHz range, while bulk viscosity is regarded as a frequency-dependent parameter~\cite{Graves1999}, resulting from the competition between the internal relaxation time of molecules and the period of acoustic waves. Therefore, the values measured at MHz frequencies should not be directly applicable to much higher frequencies, such as in light scattering experiments, where the (hypersound) acoustic waves are in the GHz domain. For example, Pan~\emph{et al.} found that bulk viscosity for CO$_2$ in their coherent Rayleigh-Brillouin scattering (CRBS) experiment, is 1000 times smaller than the sound absorption value~\cite{Pan2005}. They suggested that values of bulk viscosity at high frequencies could be derived by comparing the light scattering profiles of gases to accurate models developed by Boley \emph{et al.}~\cite{Boley1972} and Tenti \emph{et al.}~\cite{Tenti1974}, given that in these models the only unknown parameter is $\eta_b$.

In this Letter, we present measurements of spontaneous Rayleigh-Brillouin scattering (SRBS) profiles of N$_2$ in a temperature range of 255 K to 340 K and a pressure range of 850 mbar to 3400 mbar. The measured scattering profiles are compared to the so-called Tenti S6 model~\cite{Tenti1974}, which is generally considered as the most accurate model to describe the RB-scattering profile~\cite{Young1983}. Implicit in the model is that the Brillouin side peaks to the central Rayleigh peak in the scattering profile are shifted by~\cite{Boyd2008}:
\begin{equation}
\label{equ:brillouin-freq}
 \Omega_B = \pm 2n \omega \frac{v}{c} \sin \frac{\theta}{2}
\end{equation}
with $n$ and $v$ the index of refraction and the sound velocity in the gas, and $\omega$ and $\theta$ the angular frequency of the light and the scattering angle. The Brillouin side peaks exhibit a profile, associated with the damping of acoustic waves, and dependent of the thermodynamic properties of the gaseous medium as well as the light scattering parameters, yielding a Lorentzian profile of full width half maximum:
\begin{equation}
\label{equ:brillouin-width}
 \Gamma_B = \frac{1}{\rho v^2} \left[ \frac{4}{3}  \eta_s + \eta_b +  \frac{\kappa}{C_p} (\gamma - 1) \right]
  \Omega_B^2
\end{equation}
with $\rho$ the density, $\kappa$ the thermal conductivity, and $\gamma=C_p/C_v$.
The code implementing the Tenti model (version S6) was based on that of Pan~\emph{et al.}~\cite{Pan2004}, and was used for previous studies on spontaneous and coherent RB scattering in gases~\cite{Vieitez2010,Witschas2010}. This method via the Tenti model must be followed for extracting $\eta_b$ in gases where the central Rayleigh peak overlays the Brillouin side peak, unlike for liquids where the Brillouin features are fully isolated and $\eta_b$ can be determined directly by measuring the width $\Gamma_B$~\cite{Xu2003}.

\begin{figure*}[htb]
\centerline{\includegraphics[width=16cm]{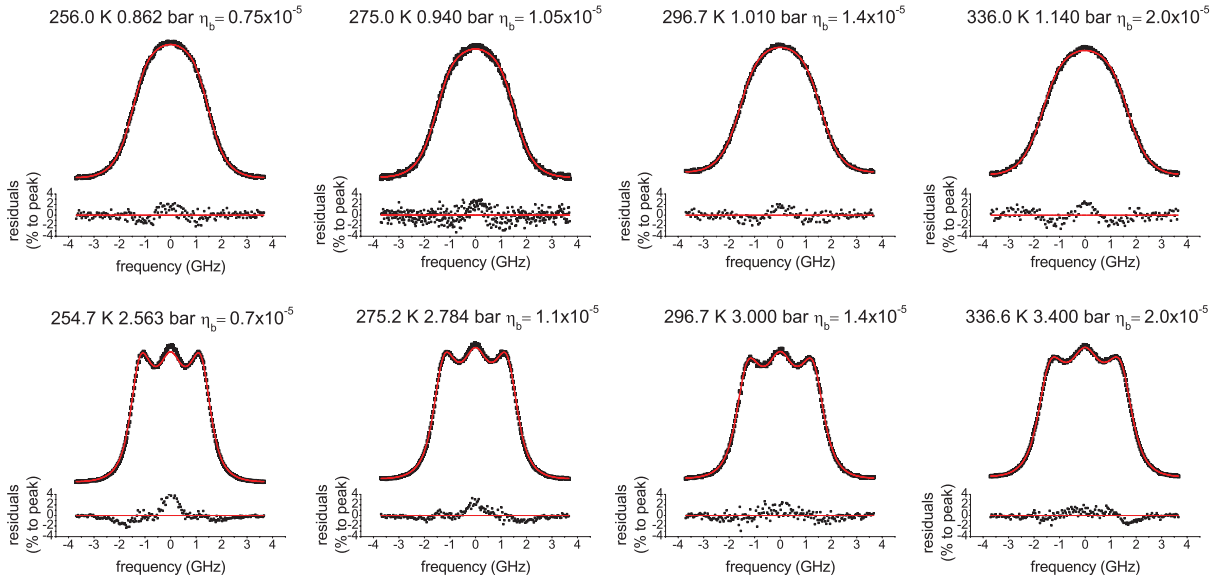}}
\caption{\label{fig:All_Spectra}Rayleigh-Brillouin scattering profiles (black dots) as measured for various ($p$, $T$) pressure-temperature combinations as specified. A comparison is made with calculations via the  Tenti S6 model (red lines), convolved for the instrument width of 232 MHz and for values of the bulk viscosity $\eta_b$, deduced from the profiles recorded at 3 bar. Residuals between the measurements and the calculations are given underneath.}
\end{figure*}

Details of the experimental setup and methods for measuring high signal-to-noise RB scattering profiles have been reported in~\cite{Gu2012}. The profiles are recorded for scattering at $\theta=90^{\circ}$ induced by an effective intra-cavity circulating power of 5 Watt at $\lambda=366.8$ nm, via a plano-concave Fabry-Perot interferometer (FPI) with an instrument linewidth of 232 MHz. For each measurement, the scattering cell is initially charged to one of the designated pressures, namely 1 bar or 3 bar, at room temperature, followed by sealing the cell and then setting the temperature to one of the designated values: 255 K, 275 K, 297 K or 336 K. The actual pressure of each measurement thus differs from the initial pressure, while the number density of the gas molecules remains the same. The actual pressure is derived via the ideal gas law.

Scattering profiles of N$_2$ at eight different ($p$, $T$) pressure-temperature combinations are shown as black dots in Fig~\ref{fig:All_Spectra}. Since the effect of $\eta_b$ is most significant at the highest pressures, where the Brillouin side peaks become pronounced (see Fig.~\ref{fig:All_Spectra}), the data recorded with an initial pressure of 3 bar are used for determining $\eta_b$.

\begin{figure}[htb]
\centerline{\includegraphics[width=6cm]{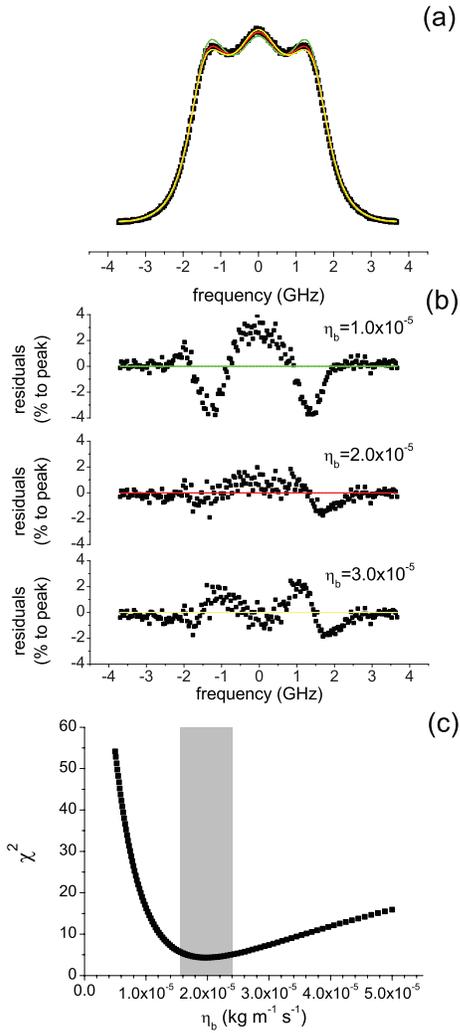}}
\caption{ \label{fig:N2_bulk_scan_with_residuals} (a) Experimental Rayleigh-Brillouin scattering profile in N$_2$ for 3.40 bar and 336.6 K (black dots), and convolved Tenti S6 calculations for bulk viscosity being 1.0$\times 10^{-5}$ (green line), 2.0$\times 10^{-5}$ (red line) and 3.0$\times 10^{-5}$ (yellow line) kgm$^{-1}$s$^{-1}$, respectively. (b) Residuals between measured and calculated scattering profiles for these three values of the bulk viscosity. (c) A plot of the $\chi^2$ as a function of bulk viscosity. The optimized value of bulk viscosity is found at the minimum of $\chi^2$, with the gray area indicating the estimated statistical error, calculated according to procedures discussed in~\cite{Meijer2010,Vieitez2010}.}
\end{figure}

Fig.~\ref{fig:N2_bulk_scan_with_residuals} elucidates the method for extracting a value for  $\eta_b$ in the comparison of the Tenti S6 model with the RB-profiles for the specific recording of an RB-profile in N$_2$ under conditions $T= 336.6$ K and $p=3.40$ bar. Panel (a) shows the measurement (black dots) and the modeled scattering profiles, for three different values of bulk viscosity, and for values of the N$_2$ transport coefficients as obtained from literature (listed in Table~\ref{tab:transport coefficients}). For the dimensionless internal specific heat capacity of internal degrees of freedom $c_{int}$ a value of 1 is used throughout. Residuals between the measurement and the three modeled scattering profiles are shown in (b). Panel (c) shows a $\chi^2$-calculation as a function of bulk viscosity employed in the Tenti S6 model.

\begin{table}
  \centering
  \caption{\label{tab:transport coefficients}Transport coefficients used for modeling the RB-profiles of N$_2$. Values for $\eta_s$ and $\kappa$ are calculated according to the Sutherland formula in~\cite{White1991}, and $\eta_b$ from the present experiment.}
  \begin{tabular}{cccc} \\ \hline
    $T$ & $\eta_s$ & $\kappa$ & $\eta_b$  \\
    K  &kg$\cdot$m$^{-1}$$\cdot$s$^{-1}$ & W$\cdot$K$^{-1}$ $\cdot$m$^{-1}$ &  kg$\cdot$m$^{-1}$$\cdot$s$^{-1}$\\ \hline
    254.7 & 1.57$\times$10$^{-5}$  & 2.28 $\times$10$^{-2}$ & 0.7$\times$10$^{-5}$ \\
    275.2 & 1.67$\times$10$^{-5}$  & 2.44 $\times$10$^{-2}$ & 1.1$\times$10$^{-5}$ \\
    296.7 & 1.76$\times$10$^{-5}$  & 2.52 $\times$10$^{-2}$ & 1.4$\times$10$^{-5}$  \\
    336.6 & 1.95$\times$10$^{-5}$  & 2.88 $\times$10$^{-2}$ & 2.0$\times$10$^{-5}$  \\
    \hline
  \end{tabular}
\end{table}

\begin{figure}[htb]
\centerline{\includegraphics[width=8cm]{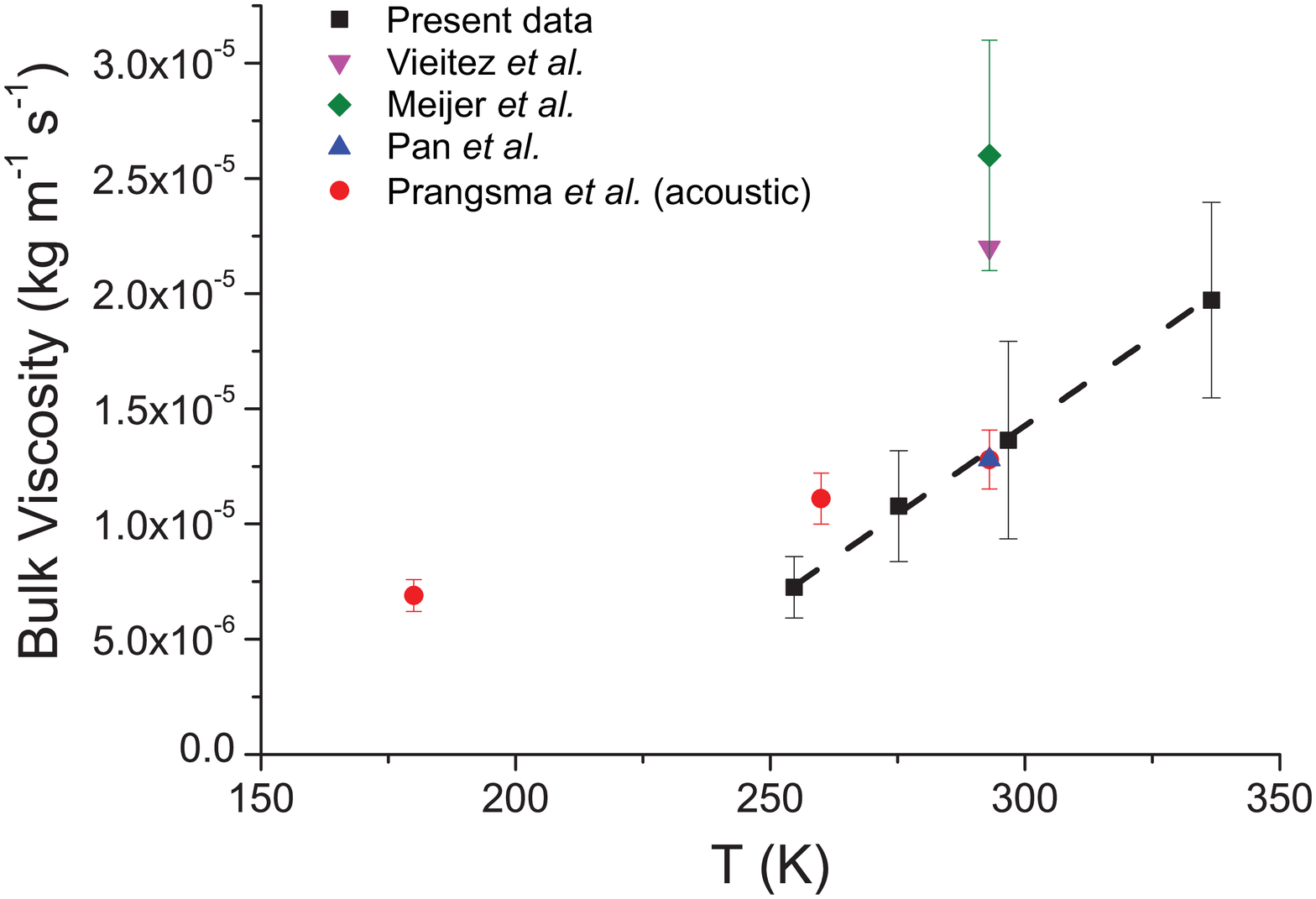}}
\caption{ \label{fig:Bulk_viscosity_comparison} Comparison of bulk viscosity measured from different experiments. Note that the result of Pan~\emph{et al.}~\cite{Pan2004} overlays a data point by Prangsma~\emph{et al.}~\cite{Prangsma1973}. Data of Vieitez~\emph{et al.}~\cite{Vieitez2010} and Meijer~\emph{et al.}~\cite{Meijer2010} also included.}
\end{figure}

This procedure of optimizing $\eta_b$ was applied to the RB-scattering measurements for initial pressure of 3 bar N$_2$. The resulting values for $\eta_b$ and their uncertainties are plotted in Fig.~\ref{fig:Bulk_viscosity_comparison}, combined with values from literature.  Prangsma \emph{et al.}~\cite{Prangsma1973} determined bulk viscosities for N$_2$ using sound absorption measurements in the temperature range $T=70 \sim 300$ K. The experiment investigated a wide range of acoustic frequency to pressure ratios, but all in the MHz domain.
Pan \emph{et al.}~\cite{Pan2004} used the value from Prangsma \emph{et al.}~\cite{Prangsma1973} and found good agreement between their CRBS profile and a calculation using the Tenti model (the S7 variant)~\cite{Boley1972}, suggesting that the value of bulk viscosity for N$_2$ obtained at MHz frequencies is also valid for the GHz range.
Cornella \emph{et al.}~\cite{Cornella2012} successfully modeled CRBS profiles in N$_2$ assuming a $\eta_b/\eta_s$ ratio of 0.73 from \cite{Prangsma1973}, valid at room temperature, and extrapolated this to 500 K.
Values previously obtained by Vietez \emph{et al.}~\cite{Vieitez2010} using SRBS at 3 bar N$_2$ slightly deviate; however, no uncertainty was specified and if a similar uncertainty is assumed as in the present study agreement within combined $1\sigma$ follows. Meijer \emph{et al.}~\cite{Meijer2010} using CRBS (at 532 nm) at 5 bar N$_2$ deduce an even larger value, but still agreement within $2\sigma$ results.

The present experimental results for $\eta_b$ in the temperature interval $254 - 337$ K, shown as black dots in Fig.~\ref{fig:Bulk_viscosity_comparison} show a linear dependence with temperature, roughly similar as in~\cite{Prangsma1973}. While the data of~\cite{Prangsma1973} extend to temperatures as low as 180 K, and the present data extend to 337 K, for the overlapping range $250 -300$ K good agreement is found. It is assumed that for dilute gases the bulk viscosity is independent of pressure, similar to shear viscosity and thermal conductivity~\cite{White1991}. The RB-profiles recorded for 1 bar N$_2$ gas, shown in the upper panels of Fig.~\ref{fig:All_Spectra}, are modeled with the $\eta_b(T)$ values obtained for 3 bar, also yielding good agreement.
While the shear viscosity $\eta_s$ is known to exhibit a linear temperature dependence in the window $254 - 337$ K~\cite{White1991}, the ratio $\eta_b/\eta_s$ grows from 0.46 to 1.01 for the present data. This behaviour may be related to the freezing out of internal degrees of freedom at lower temperatures.

A general conclusion is drawn that for pure nitrogen (N$_2$) gas the bulk viscosity at acoustic frequencies in the MHz regime is the same as for hypersound frequencies in the GHz regime. This result is surprising in view of the results in carbon dioxide (CO$_2$) gas where differences by orders of magnitude were found~\cite{Pan2005}.

This work has been supported by the European Space Agency contract number 21369 under the supervision of Anne Grete Straume and Olivier Le Rille. The code for computing the Tenti S6 model was obtained from Xingguo Pan and modified by Willem van de Water.


\end{document}